\documentclass[aps,prl,superscriptaddress,reprint]{revtex4-1}
\usepackage{graphicx}
\usepackage[permil]{overpic}
\usepackage{latexsym}
\usepackage{amsmath}
\usepackage{amssymb}
\usepackage{upgreek}
\usepackage{xcolor}
\usepackage[colorlinks]{hyperref}
\hypersetup{allcolors=blue}
\begin{document}
\title{The Stone-Wales trefoil: a new topological defect in graphene.}
\author{M. V. Kondrin}
\email{mkondrin@hppi.troitsk.ru}
\affiliation{Institute for High Pressure Physics RAS, 108840 Troitsk, Moscow, Russia}
\author{Y.B. Lebed}
\affiliation{Institute for Nuclear Research RAS,  Moscow, Russia}
\author{V.V. Brazhkin}
\affiliation{Institute for High Pressure Physics RAS, 108840 Troitsk, Moscow, Russia}
\begin{abstract}
Specific heat of graphite at high temperatures demonstrates significant deviation from the ``3R'' Dulong-Petite law. It can be explained by accumulation of defects with formation energy as low as 1.45 eV. Based on the density functional theory calculations, we propose a possible candidate of topological defects, whose energy is comparable to the growth activation energy.
\end{abstract}
\maketitle

\begin{figure}
\includegraphics[width = \columnwidth]{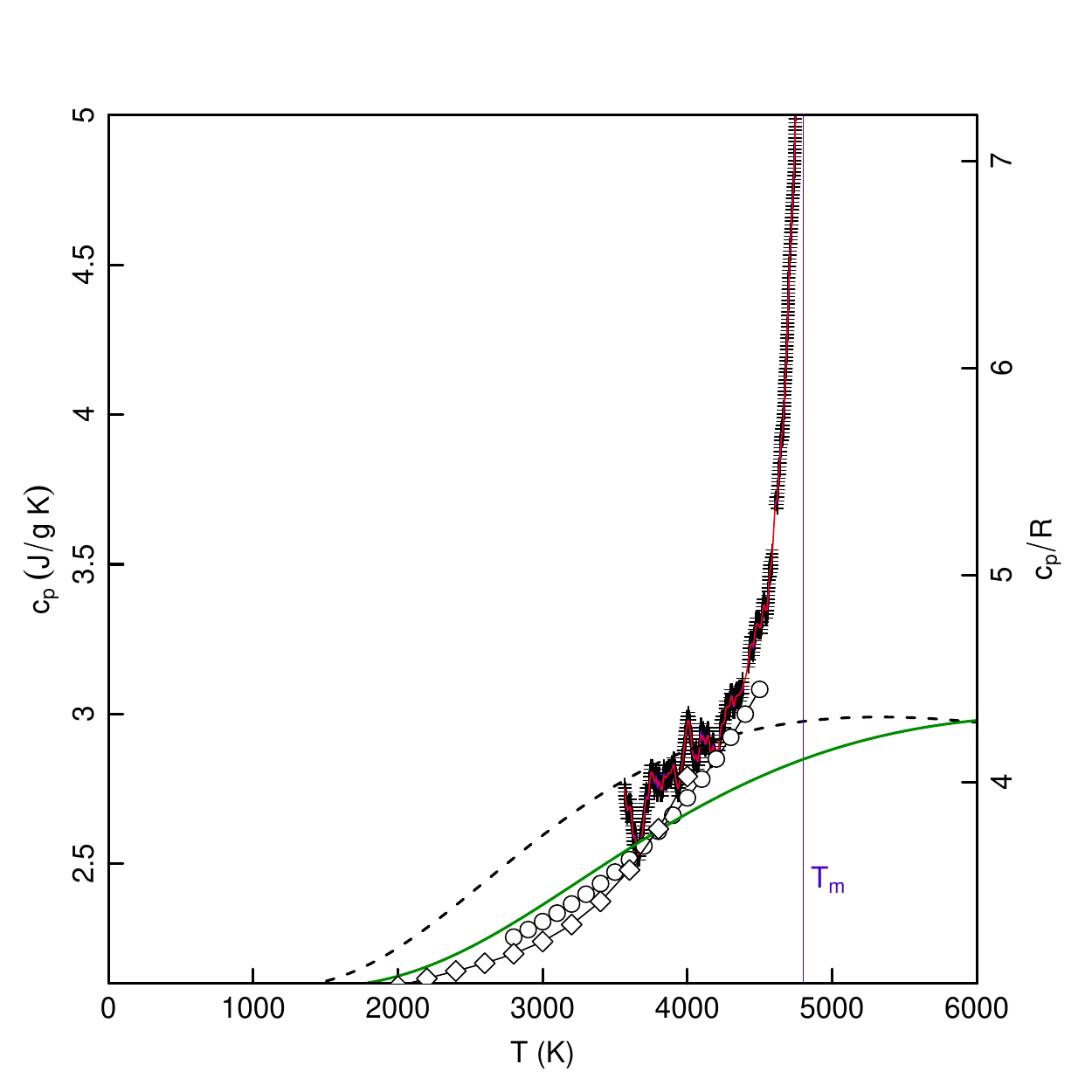}
\caption{Excessive specific heat in graphite. Symbols stand for experimental data ($\diamond$-\cite{buchnev:ht87},$\circ$-\cite{sheindlin:spd88},$+$-\cite{savva:c16}), curves depict Eq.\ref{f1} with different $E_a$ (dashed line -- $E_a = 1.16$ eV, solid -- $1.45$ eV) and $Q_a=3$. $T_m$ marks melting temperature of graphite according to experimental data\cite{savva:c16,savva:pu20}. }
\label{f:0} 
\end{figure}

\begin{figure}
\includegraphics[width = \columnwidth]{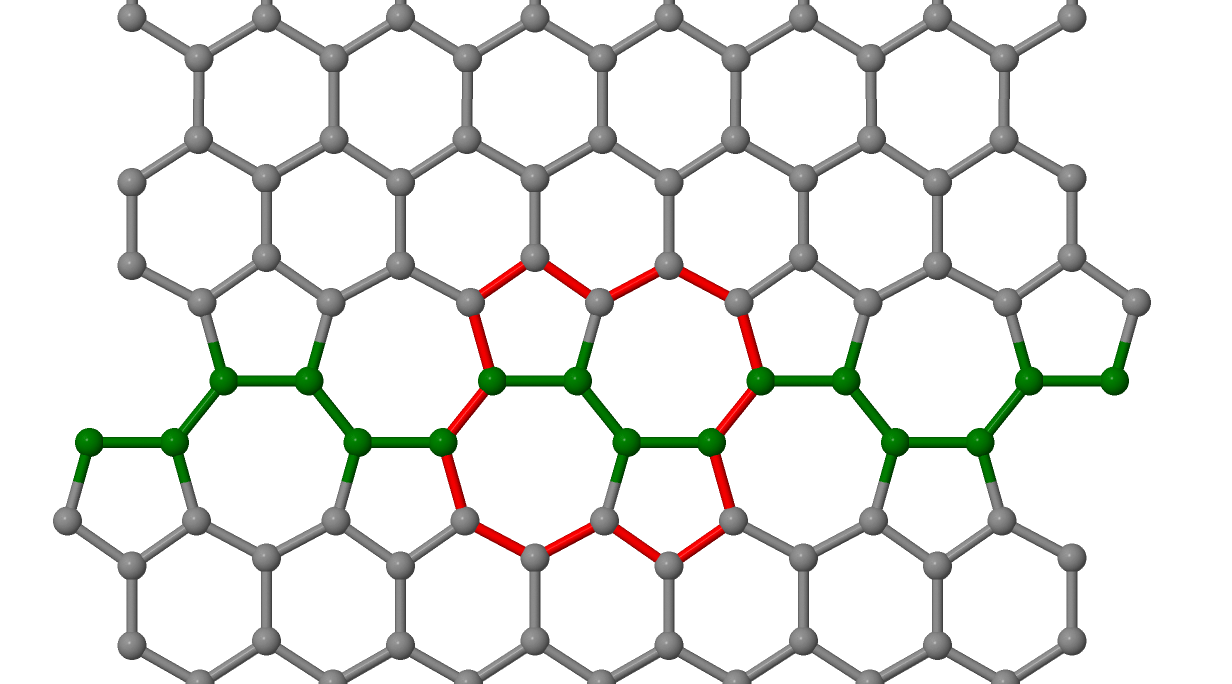}
\caption{Single extended defect (marked in green) in graphene obtained by reconstruction of interatomic bonds with formation of 5- and 7-membered cycles. It can be regarded as concatenation of individual Stone-Wales defects (one of which is outlined in red).}
\label{f:1} 
\end{figure}

The isolated Stone-Wales defect is a well-studied defect in fullerene, graphene and nanotubes \cite{tiwari:s23}. However, these defects are too rare  (because of  high formation energy, more than 4.2 eV \cite{ertekin:prb09}) 
 to have significant impact on  thermodynamic properties of graphene, so the  deviation of high temperature heat capacity from the Dulong-Petite law, which was observed in experiments on melting of graphite\cite{buchnev:ht87,sheindlin:spd88,savva:c16,savva:pu20}(see Fig.~\ref{f:0}) cannot be explained by their accumulation.  It was  demonstrated for graphite , that the difference $c_P - c_V = T (\alpha)^2 V_0 B_T $ ($\alpha= 3.2$ 10$^{-5}$ K$^{-1}$ -- thermal expansion coefficient, $B_T$ = 34 GPa bulk modulus and $V_0=$ 6 10$^{-6}$ m$^{-3}$/mol -- specific volume ) is negligible in the solid phase \cite{ivanov:prb26}. When the Stone-Wales defects are concatenated and form a string along one of crystallographic directions (see Fig.~\ref{f:1}) , much smaller formation energy than that of individual  defect is needed.   This kind of extended defect relates to the crystal  structure of carbon, named $\psi$-graphene \cite{csanyi:prb07,li:jpcl17}, with extremely low additional energy compared to pure graphene. Such a stacking 
 fault of graphene was investigated by Ariza using  the method of molecular dynamics  {\em et al.} \cite{ariza:pm12}. The Stone-Wales defects included into this extended defect  have significantly smaller formation energy, equal to 2.32 eV \cite{kondrin:prl21}. Drastic decrease in the formation energy is probably caused by full relaxation of strains present in the Stone-Wales defect when it is directly embedded into graphene lattice. Still, application of this kind of extended defect to describe the excess  of specific heat encounters some difficulties.

\begin{figure}
\begin{overpic}[width = 0.49\textwidth]{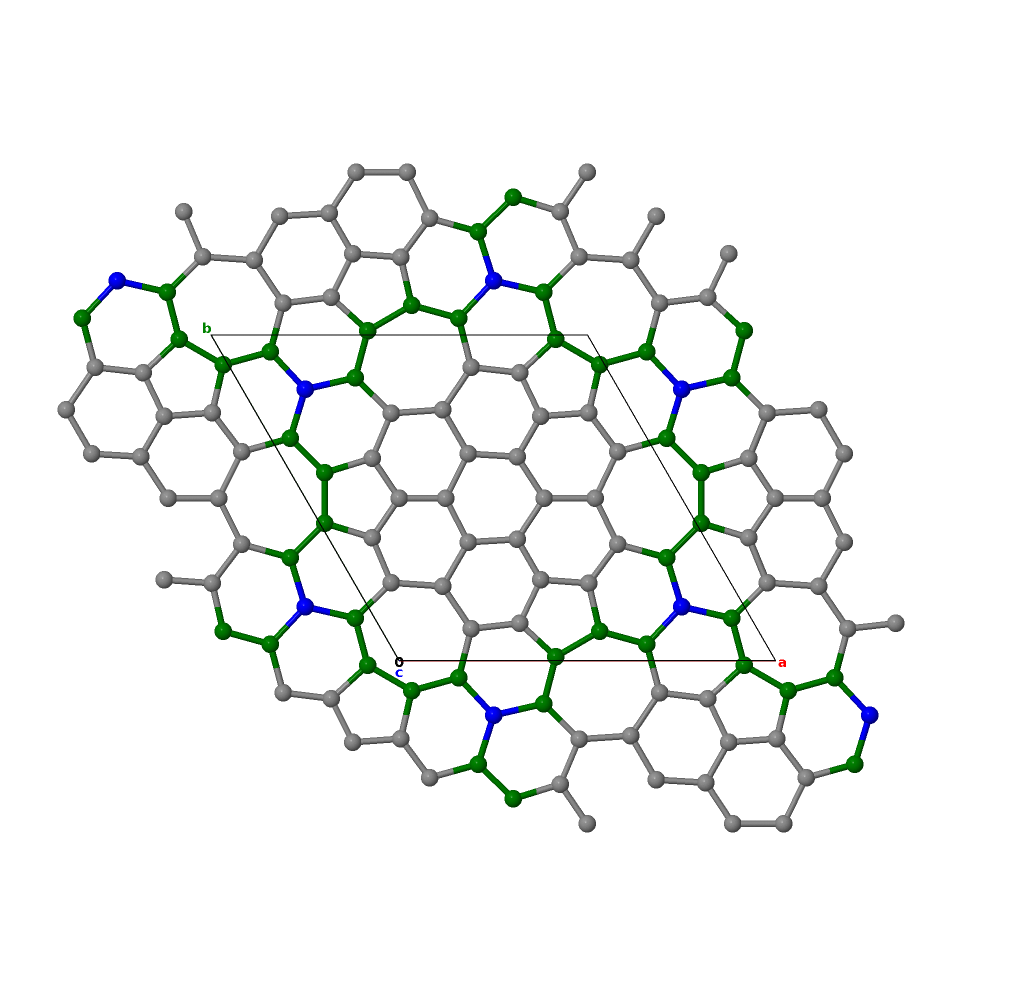}
\put(10,90){\fcolorbox{black}{white}{\Large a)}}
\end{overpic}
\begin{overpic}[width = 0.49\textwidth]{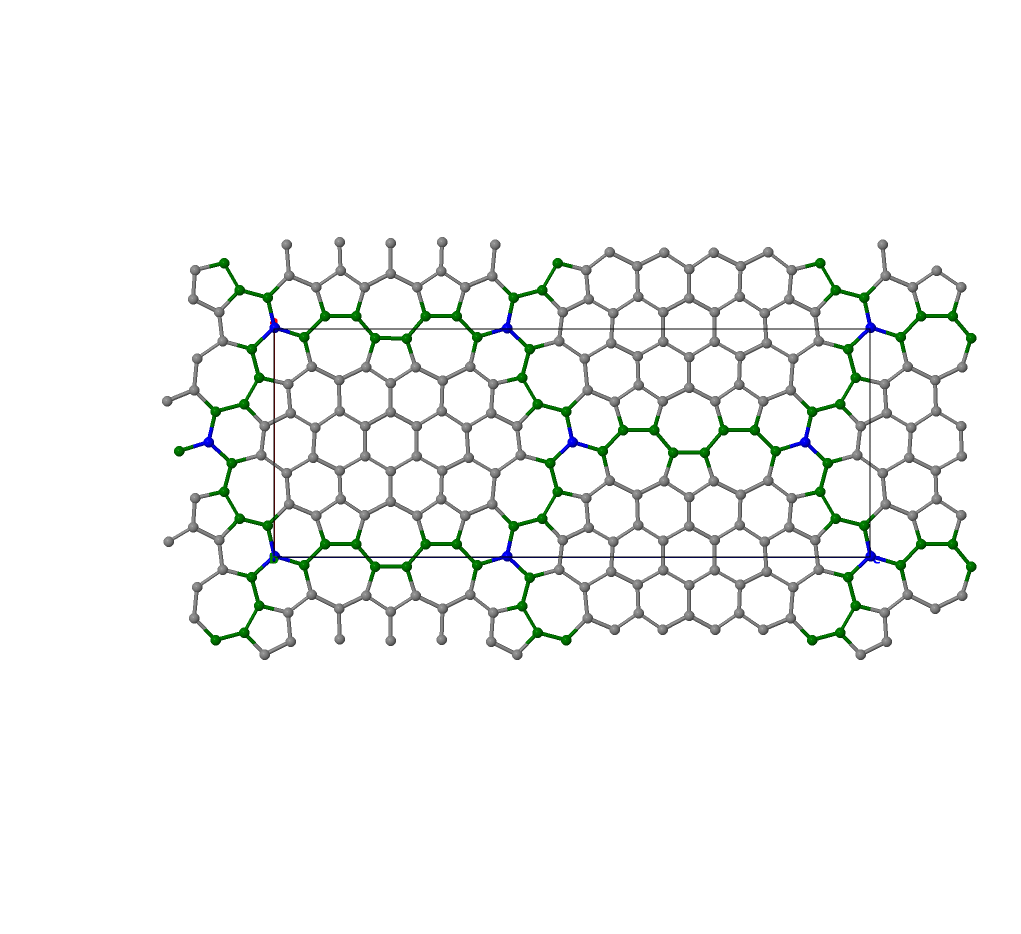}
\put(10,90){\fcolorbox{black}{white}{\Large b)}}
\end{overpic}

\caption{Regular models of crossing of extended defects with different concentrations\cite{kondrin:drm24}. Green bonds depict network of defects , blue balls mark the point where the three defects meet, gray bonds connect area of unperturbed graphene. }
\label{f:2}
\end{figure}

Assuming activation law for the defect concentration the {\em point} defects contribute to specific heat as \cite{kondrin:prl21}:
\begin{equation}
 \Delta c_V=Q_a\frac{E_a^2}{k_B T^2} \exp\left(-\frac{E_a}{k_BT}\right)
\label{f1}
\end{equation} 
where $E_a$ -- formation energy of point defect and $Q_a$--configurational multiplier related to activation entropy $S_a=k_B\log(Q_a)$. This function has a maximum approximately equal to $ 0.45 k_B Q_a$, so by comparison of actual $\Delta c_V$ observed in experiment with $0.45 R$ one can roughly estimate the lower limit on $Q_a$.  For graphene this limit is approximately equal to 3. This value can be compared to the number of extended defects in which an atom can participate. This means that in case that the excess of specific heat is caused by extended defects,  they should cross. Crystallographic model of such a crossing was proposed lately \cite{kondrin:drm24} and is depicted in Fig.~\ref{f:2}. These structures can be referred to as irregular networks of extended defects and   they can form  in 2D crystals at the melting temperature \cite{kondrin:pccp26,kondrin:jetpl26}. On the other hand, application of Eq.~\ref{f1} with $Q_a=3$ to describe excess specific heat of graphite yields the formation energy $E_a \approx 1.16$ eV, that is growth of extended defects occurs by addition of individual dislocations (halves of the Stone-Wales defect) rather then of whole Stone-Wales defect. In our previous works \cite{kondrin:prl21,kondrin:drm24} we erroneously assumed that extended defect in the solid graphene grows by additions of single atoms, so we overlooked this problem. Although dissociated Stone-Wales defects are considered in the literature (see e.g. \cite{ertekin:prb09}) they always appear in pairs. Beside that, dissociation of the Stone-Wales defect is accompanied by energy penalty and formation energy of dissociated dislocation dipole turns out to be greater than 6.2 eV \cite{ertekin:prb09}.

Usually the deviation of high-temperature specific heat from the Dulong-Petite law is attributed to anharmonism, so one may think that defects are alternative mechanism to anharmonicity. However, as we demonstrated earlier \cite{kondrin:pssb22,kondrin:jetp23} defects (especially extended ones) are another way to describe anharmonic effects. Indeed, stacking fault depicted in Fig.~\ref{f:1} as it was shown by Ariza {\em et al.} \cite{ariza:pm12} is created in response to shear strain, produced when one half plane is shifted along another one by half a period. Such a strong strain leads to breaking of bonds between atoms and their subsequent reconstruction in more energetically favorable pattern. It may be thought of as if the potential energy surface in coordinates Energy-Shift has well defined local minimum corresponding to this extended defect that is it can be regarded as strong anharmonicity. Such a strong anharmonicity poorly can be described by perturbative methods which is an essence of anharmonic approach but it perfectly fits into framework of defect model.

In the following we will show that such an isolated defect does exist in graphene. This sort of defect is dynamic, that is, it grows with temperature rising, and the growth of this defect is an activational process with activation energy $E_a \approx 1.45$ eV and $Q_a = 3$.

For calculation of graphene defects  formation energies the  QuantumESPRESSO software package  was used \cite{giannozzi:jopcm17}. For the density functional calculation we employed the Perdew-Burke-Ernzerhof exchange correlation method with norm-conserving pseudopotentials with the energy cutoff 70 Ry.  For integration over Brillouin zone $\Gamma$ only point was used. The crystal lattices and atom positions at fixed external pressure were fully optimized with dimension normal to the carbon layer being kept fixed at 10 \AA\, until residual force on every atom did not exceed 0.001 Ry/bohr and additional stress -- 0.5 kbar. The formation energy of the defect was calculated by taking the energy difference between defected structure ($E$) and  defect-free diamond  lattice ($E_{host}$) of  similar dimensions:
\begin{equation*} 
\Delta E = E-E_{host} n/n_{host}
\end{equation*} 
where $n$ and $n_{host}$ are the number of carbon atoms in defected and pristine graphite crystals. For preparation of input files we heavily used \verb|cif2cell| \cite{cif2cell} program.

\begin{figure}
\includegraphics[width = \columnwidth]{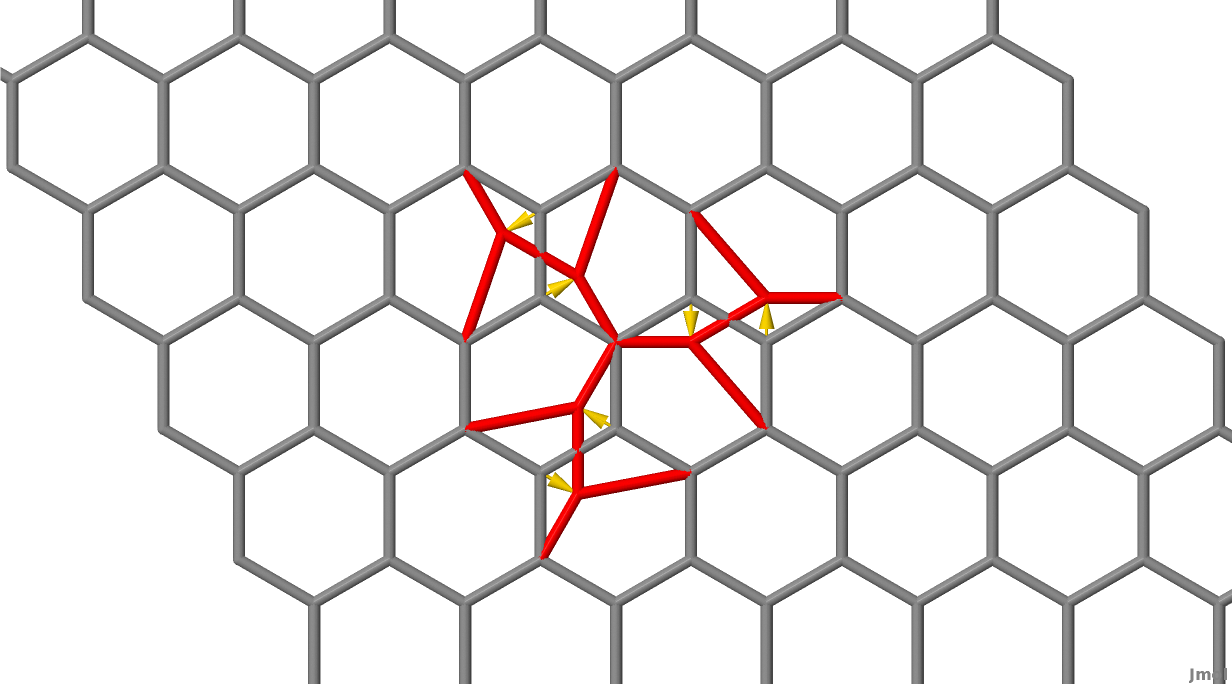}
\caption{Reconstruction of graphene bonds leading to formation of 1-1-1 Stone-Wales trefoil (shown in red). Gray bonds mark nearly unperturbed graphene. Yellow arrows show movements of carbon atoms in the process of bond reconstruction.     }
\label{f:3} 

\end{figure}

\begin{figure}
\begin{overpic}[width = 0.49\textwidth]{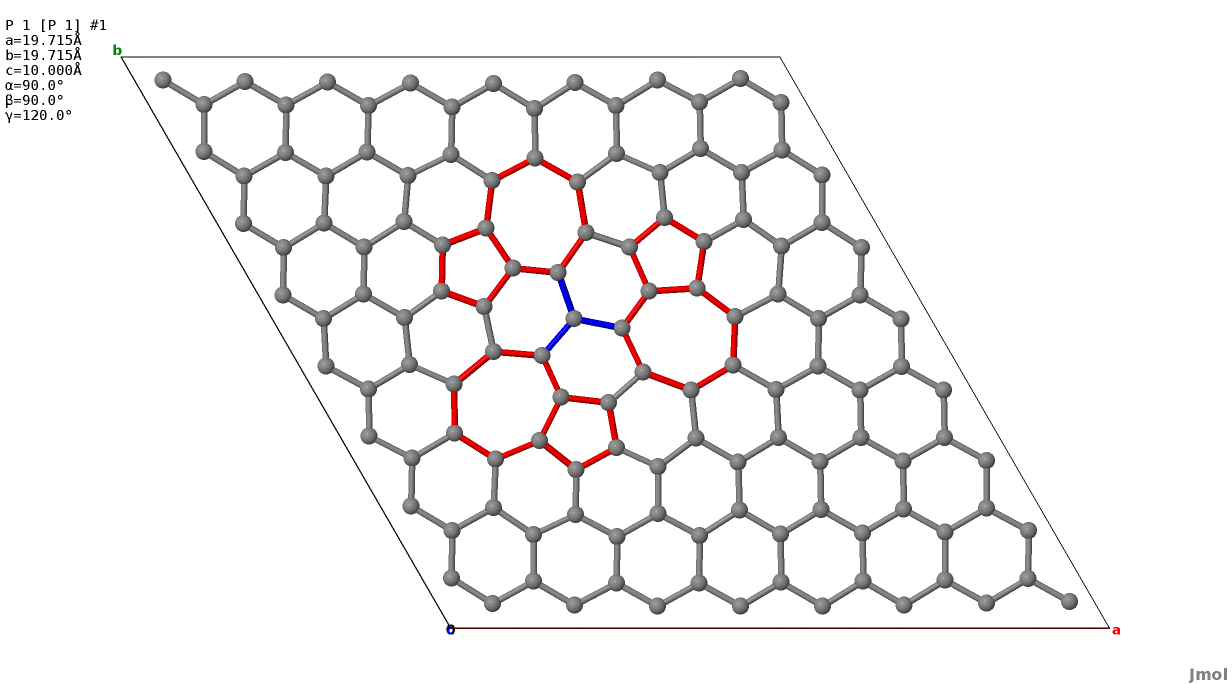}
\put(10,90){\fcolorbox{black}{white}{\Large a)}}
\end{overpic}
\begin{overpic}[width = 0.49\textwidth]{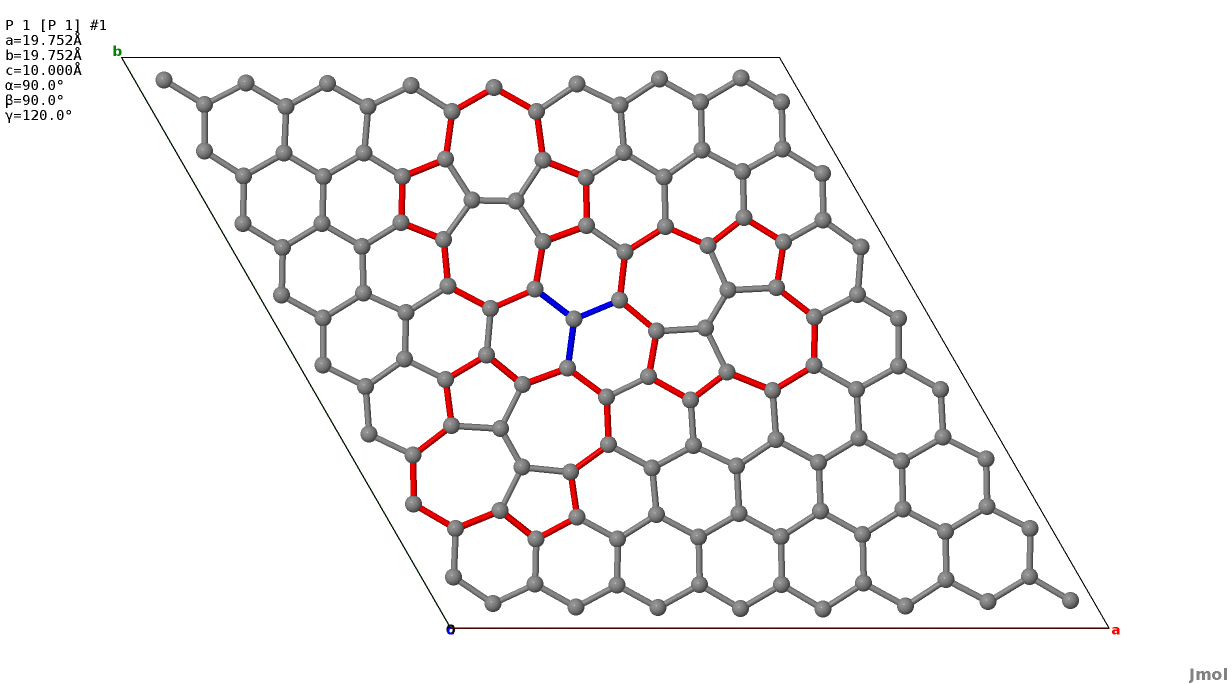}
\put(10,90){\fcolorbox{black}{white}{\Large b)}}
\end{overpic}
\begin{overpic}[width = 0.49\textwidth]{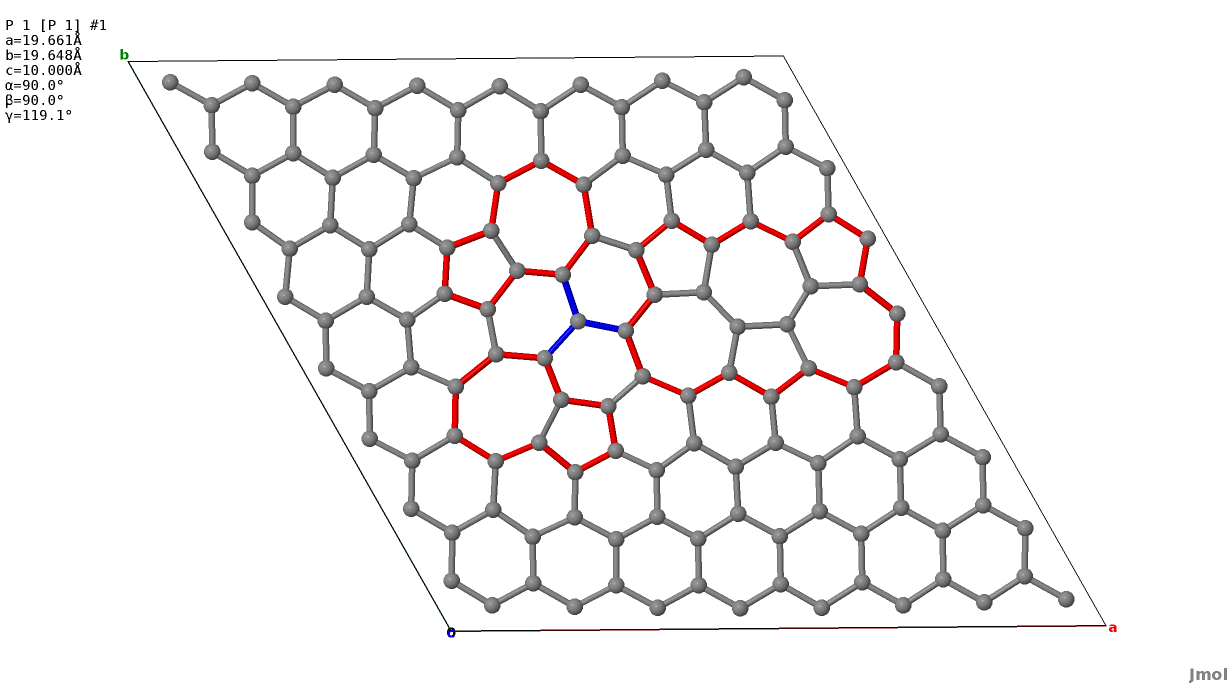}
\put(10,90){\fcolorbox{black}{white}{\Large c)}}
\end{overpic}

\caption{Relaxed structures of Stone-Wales trefoils. a) 1-1-1, b) 2-2-2, c) 1-1-3. Blue bonds designate stem of trefoil, red bonds outline leaves. }
\label{f:4}
\end{figure}

The idea of a defect named the Stone-Wales trefoil is based on consideration of individual node of networks depicted in Fig.~\ref{f:2}. Bond reconstruction of one node embedded in unperturbed graphene is shown in Fig.~\ref{f:3}. It involves rotation as a whole of three bonds which meet in the node (``stem'') and flipping of bonds next to them in the manner characteristic of isolated Stone-Wales defect, so the resulting structure contains one dislocation in every ``leaves'' of defect. We will name it 1-1-1 Stone-Wales trefoil. The stem of trefoil consists of hexagons, since 5- and 7-membered cycles produced during flipping of individual bonds is self-healed when three bonds flip at once. The resulting structure can be subsequently relaxed. The optimized structure is shown in Fig.~\ref{f:4} a).

Previous consideration is applicable only to defects which comprise an odd number of dislocations (we will call such defect an odd Stone-Wales trefoil). The different flavor of defect is even trefoils (like 2-2-2 one shown in Fig.~\ref{f:4} b). Their difference from odd trefoils is that their stem is not rotated and oriented as in unperturbed graphene (compare orientation of stems in Fig.~\ref{f:4} a) and b)). So growth of trefoils involves a) rotation of stems (which change the number of dislocations by 3) and b) addition of Stone-Wales defects to the outer part of the leaves (which change the number of dislocations by even number). For example addition of one Stone-Wales defect to 1-1-1 trefoil results in creation of 1-1-3 trefoil shown in Fig.~\ref{f:4} c). These two operations let one construct a sequence of trefoils which grows by addition of one dislocation starting from 1-1-1 trefoil -- 1-1-1, 0-2-2, 1-1-3, 2-2-2, 1-3-3, 2-2-4, 3-3-3 {\em etc}. How it looks like in energy?

For this purpose we calculated the formation energy in the diluted limit of four trefoils embedded in 128 atoms supercell (see Fig.~\ref{f:4}) -- 1-1-1 ($E_f=$9.29 eV), 1-1-3 ($E_f=$12.77 eV), 2-2-2 ($E_f=$12.73 eV) and 3-3-3 ($E_f=$20.33 eV). Formation energy $E_f$ is a sum of two inputs -- formation energy of stem ($E_{stem}$ we consider it to be different for even and odd trefoils) and formation energy of $n$ dislocations $E_d$ $E_f=E_{stem}+nE_d$. This yields $E_d \approx 1.8$ eV, $E_{stem}^{even} \approx 1.9$ eV,  $E_{stem}^{odd} \approx 3.9$ eV. In agreement with that was written above  unrelaxed strain present in Stone-Wales defects embedded in graphene lattice increases formation energy of dislocations more than one and a half times.

\begin{figure}
\includegraphics[width = \columnwidth]{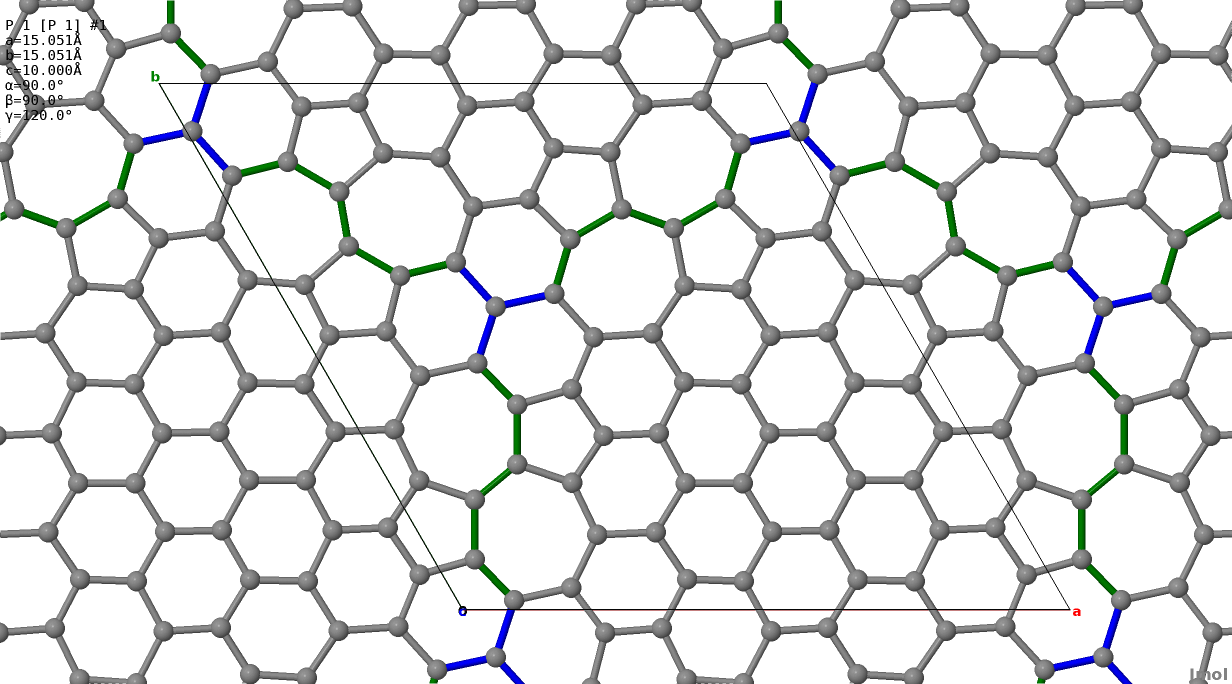}
\caption{2-2-2 Stone-Wales trefoil as part of network of extended defects.}
\label{f:5} 
\end{figure}

However, diluted limit is a bad approximation for high-temperature thermodynamic properties of graphene, where concentration of defects should be high. So, we consider the formation energy of trefoils when they are taking part  of irregular network of extended defects. This would provide defect energy in the saturated limit. For networks consisting of odd trefoils, we take the data published earlier \cite{kondrin:drm24} (see Fig.~\ref{f:2}) and for network consisting of even trefoils we calculate the energy of structure depicted in  Fig.~\ref{f:5} (this structure is derived from graphene by linear transformation of axes $\textbf{a'}=3\textbf{a}+7\textbf{b};\textbf{b'}=4\textbf{a}-3\textbf{b}$.). For structure shown in Fig.~\ref{f:2} a) (3 dislocations, 2 stems) formation energy $E_f=6.42$ eV, Fig.~\ref{f:2} b) (10 dislocations, 4 stems)  $E_f=17.16$ eV, Fig.~\ref{f:5} (6 dislocations, 2 stems)  $E_f=8.71$ eV. This yields   $E_d \approx 1.08$ eV, $E_{stem}^{even} \approx 1.11$ eV,  $E_{stem}^{odd} \approx 1.59$ eV. The formation energy of dislocation is approximately equal to that of extended defect depicted in Fig.~\ref{f:1}. Taking the average of diluted and saturated limits we obtain for intermediate case $E_d=1.44$ eV. This is equal to the activation energy of specific heat obtained in experiments and shown in Fig.~\ref{f:0}.

It is interesting to compare formation energy of structure depicted in Fig.~\ref{f:5} with $Cmmm$ structure of $\sqrt{8} \times \sqrt{8}$ Stone-Wales graphene (periodic arrangement of the Stone-Wales defect in graphene lattice)\cite{yin:prb19}. The formation energy per Stone-Wales defect in the latter case is equal to 4.2 eV, which is substantially higher than that of three Stone-Wales defects arranged in the network of extended defects (2.9 eV, here we do not differentiate energy inputs between dislocations and stems). For odd trefoils this characteristic is higher due to higher input from stems (3.4 eV for structure depicted in Fig.~\ref{f:2} b)  but still less than that of the Stone-Wales graphene. This indicates that arrangement in network of extended defects partially compensates the delocalized stresses induced by Stone-Wales defects and present in graphene lattice.

To summarize, we propose a new topological defect in graphene -- the Stone-Wales trefoil consisting of 5-,6- and 7-membered carbon rings. The distinct feature of such a defect is its dynamic nature, that is, an ability to grow by increment of one dislocation (pair of 5- and 7-membered rings) in a time. Activation energy of such increment is as low as 1.08 eV. Comparison with high-temperature experimental specific heat measurements suggests that this type of a defect can be responsible for the observed deviation from the Dulong-Petit law.     

{\bf Acknowledgments.} This work was supported by the Russian Science Foundation under grant No. 24-12-00037.

%

\end{document}